\documentclass[twocolumn,amsmath,amssymb,floatfix]{revtex4}

\usepackage{bm}
\usepackage{amsmath}
\usepackage{epsf}
\newcommand{\beq}{\begin{equation}}

\newcommand{\eeq}{\end{equation}}
\newcommand{\beqa}{\begin{eqnarray}}
\newcommand{\eeqa}{\end{eqnarray}}

\newcommand{\bk}{{\bf k}} 
\newcommand{\bx}{{\bf x}}

\newcommand{\MNRAS}{Mon. Not. R. Astron. Soc.}

\newcommand{\Mobs}{M^{\rm obs}}
\newcommand{\Mtrue}{M}
\newcommand{\Mbias}{M^{\rm bias}}
\newcommand{\siglnM}{\sigma_{\ln \Mtrue}}
\begin{document}

%@twocolumnfalse\endcsname
\title{Self-Calibration of Cluster Dark Energy Studies: Observable-Mass Distribution}
\author{Marcos Lima$^{1}$ and Wayne Hu$^{2}$}
\affiliation{
$^{1}$Department of Physics, University of Chicago, Chicago IL 60637\\
$^{2}$Kavli Institute for Cosmological Physics, Department of Astronomy and Astrophysics,
and Enrico Fermi Institute, University of Chicago, Chicago IL 60637 \\}

\begin{abstract}
\baselineskip 11pt
The exponential sensitivity of
 cluster number counts 
to the properties of the dark energy implies a comparable sensitivity to not only the
mean but also the actual {\it distribution} of
an observable mass proxy given the true cluster mass.   For example a $25\%$ scatter
in mass can provide a $\sim 50\%$ change in the number counts at $z \sim 2$ for the upcoming SPT 
survey.  Uncertainty in the scatter of this amount would degrade dark energy constraints
to uninteresting levels.   Given the shape of the actual mass function, 
the properties of the distribution
may be internally monitored by the shape of the 
{\it observable} mass function.   
An arbitrary evolution of the scatter of a mass-independent Gaussian distribution 
may be self-calibrated to allow a measurement of the dark energy equation of state of
 $\sigma(w) \sim 0.1$.  External constraints on the mass {\it variance} of the
distribution that are more accurate than
$\Delta \siglnM^2 < 0.01$ at $z \sim 1$
can further improve constraints by up to a factor of 2.
More generally, 
cluster counts 
and their sample variance measured as a function of the observable 
provide internal consistency checks on the assumed form of 
  the observable-mass distribution that will protect against misinterpretation of the
dark energy constraints.
\end{abstract}
\maketitle

\section{Introduction}

It is well known that cluster counts as a function of their mass are exponentially sensitive
to the amplitude of the linear density field and hence the dark energy dependent growth
of structure.  Unfortunately the mass of a cluster is not a direct observable and their numbers
can only be counted as a function of some observable proxy for mass.   Typical proxies
include the Sunyaev-Zel'dovich flux decrement, X-ray temperature, X-ray surface brightness
or gas mass, optical galaxy richness, and the weak lensing shear.
The exponential sensitivity to mass translates into a comparable sensitivity to the
whole {\it distribution} of the observable given the mass 
not just the mean relationship. 

While scatter in the observable-mass relation is typically addressed in studies of the local
cluster abundance (e.g. \cite{Ikeetal01}), it is commonly ignored in forecasts for upcoming
high redshift surveys (e.g. \cite{HaiMohHol01,HuKra02,WanKhoHaiMay04}).  While it is true that scatter in the 
observable of a known form does little to degrade the dark energy information, uncertainties
in the distribution directly translate into uncertainties in the dark energy
inferences that
must be controlled.
 
 In this Paper we undertake a general study of the impact of uncertainty in 
 the observable-mass distribution on high
 redshift cluster counts.
 Previous work on forecasting prospects for dark energy constraints
 have examined the effect of scatter under specific and typically more restrictive assumptions.
For example, the change in the number counts, known as Eddington bias \cite{Edd13},
has
 been assessed for a fixed cut in signal-to-noise of cluster detection via
 the Sunyaev-Zel'dovich flux in a hydrodynamic simulation \cite{HolMohCarEvrLei00} and
 through modeling a constant scatter in mass \cite{BatWel03}.  However it is the uncertainty
 in the scatter, or the error in the correction of the bias, that degrades dark energy 
 constraints.  Along these lines,
 Levine et al.  \cite{LevSchWhi02} considered the marginalization of a constant scatter in the
 mass-temperature relation for clusters but with strong external priors on the dark energy
 parameters.
 
Prospects for the self-calibration of the {\it mean} observable-mass relation have been extensively
studied recently.  Self-calibration relies on the fact that
both the shape of the mass function  \cite{Hu03a}  and 
the clustering of clusters \cite{MajMoh03}
can be predicted from cosmological simulations.  Much of the information in the latter
can be extracted from the angular variance of the counts so that costly spectroscopy can
be avoided \cite{LimHu04}.   
Thus by demanding consistency between
the counts and their sample variance across the sky as a function of 
the observable mass, one
can jointly solve for the cosmology and the mean observable mass relation.   
 Here we show that the shape of the mass function 
is even more effective at monitoring the scatter in the observable-mass relation.

We begin in \S \ref{sec:distribution} with a discussion of our parameterization 
of the observable-mass
distribution and assess its implications for
the dark energy.  We examine the prospects for
self-calibration in \S \ref{sec:self} and conclude in \S \ref{sec:discussion}.

\section{Observable-Mass Distribution}
\label{sec:distribution}

The cosmological utility of cluster number counts arises from their exponential sensitivity
 to the amplitude of the linear density field.
For illustrative purposes, we will employ
a fit to simulations for the mass function or the differential comoving density of clusters
\cite{Jenetal01}
\begin{equation}
{d \bar n \over d\ln M} = 0.3 {\rho_{m} \over M} {d \ln \sigma^{-1} \over d\ln M}
        \exp[-|\ln \sigma^{-1} + 0.64|^{3.82}]\,,
\label{eqn:massfun}
\end{equation}
where $\sigma^2(M;z)\equiv \sigma^2_{R}(z)$, the linear density field
variance in a region enclosing $M=4\pi R^3\rho_m/3$ 
at the mean matter density today $\rho_{m}$.     

\begin{figure}[tb]
\centerline{\epsfxsize=3.4in\epsffile{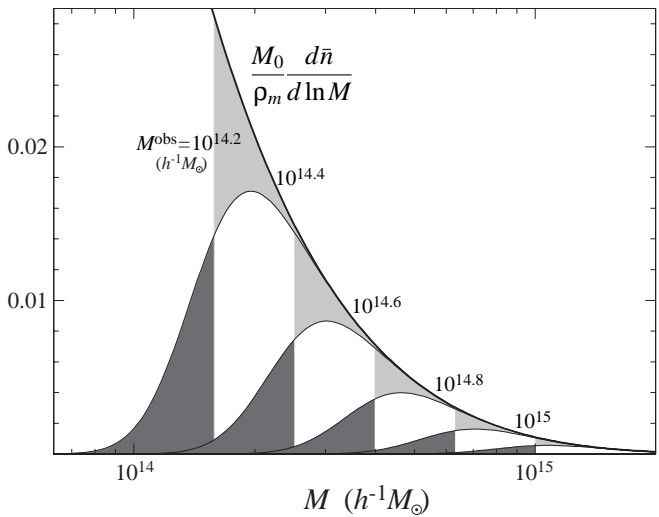}}
\caption{\footnotesize 
Scatter of $\siglnM=0.25$ in the observable-mass relation changes the  mass distribution of clusters above an observable threshold $\Mobs$ (curves) to provide an  excess of clusters scattering
up (dark shaded) versus down (light shaded) across the threshold.   Here the intrinsic mass function (thick line) has been normalized
to $M_0=10^{14} h^{-1} M_\odot$ and evaluated at $z=0$. }
\label{fig:leak}
\end{figure}

To exploit this exponential sensitivity, the observable-mass distribution must be known to
a comparable accuracy. 
Let us consider the probability of assigning a mass $\Mobs$ to a cluster of true mass $M$
to be given by a Gaussian distribution in $\ln M$ as motivated by the observed
scatter in the
scaling relations between
typical cluster observables (e.g. \cite{Ikeetal01})
\begin{equation}
p(\Mobs | \Mtrue) = {1 \over \sqrt{2\pi \siglnM^2} } \exp\left[ -x^2(\Mobs) \right] \,,
\end{equation}
where
\begin{equation}
x(\Mobs) \equiv { \ln \Mobs - \ln M - \ln  \Mbias \over \sqrt{2 \siglnM^2}} \,.
\end{equation}
For simplicity we will allow the mass variance
 $\siglnM^2$ and the mass bias $\ln \Mbias$ to vary with redshift but not mass.  
   We implicitly exclude sources of scatter due to
 noise in the measurement of $\Mobs$ which certainly would depend on $\Mobs$
 but in a way that is known given the properties of a specific survey.  
 More generally, our qualitative results will hold so long as any trend in mass at a fixed
 redshift is known.  
 
 \begin{figure}[tb]
\centerline{\epsfxsize=3.4in\epsffile{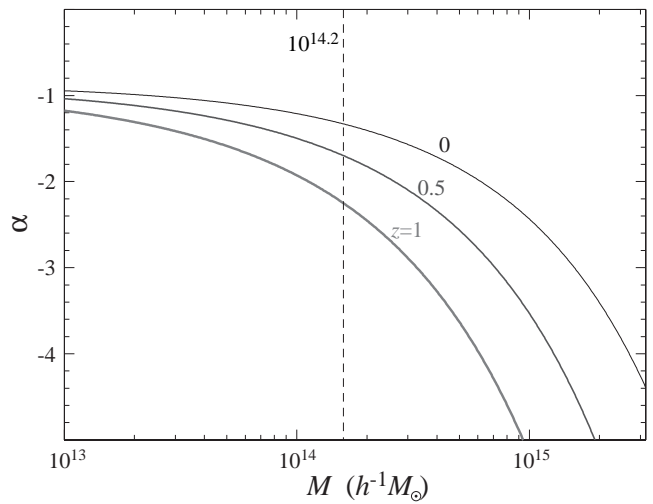}}
\caption{\footnotesize 
Local
power law index $\alpha$ of the mass function as a function of mass $d\bar n/d\ln M \propto M^{\alpha}$
 for $z=0$, $0.5$ and $1$.
The relative importance of scatter versus bias
 can be scaled through $\alpha$ and Eqn.~(\ref{eqn:alpha}) 
 to alternate mass and redshift ranges than considered here.}
\label{fig:slope}
\end{figure}

The average number density of clusters within a range defined by cuts in the observable mass
$\Mobs_i \le \Mobs \le \Mobs_{i+1}$ is
\begin{align}
\bar n_i & \equiv \int_{\Mobs_i}^{\Mobs_{i+1}} {d \Mobs \over \Mobs}
\int {dM \over M} { d \bar n \over d\ln M}
p(\Mobs | \Mtrue)\nonumber \\
& = \int {d M \over M} { d \bar n \over d\ln M} {1\over 2} \left[ {\rm erfc}(x_i) - {\rm erfc}(x_{i+1}) \right] \,,
\label{eqn:binneddensity}
\end{align}
where $x_i = x(\Mobs_i)$.  The mean number of clusters in a given volume $V_i$ is then
\begin{align}
\bar m_i = \bar n_i V_i \,.
\end{align}
Note that in the limit that $\siglnM^{2} \rightarrow 0$ and $\Mobs_{i+1} \rightarrow \infty$,
$\bar m_i$ is the usual cumulative number counts above some sharp mass threshold.

An unknown scatter or more generally uncertainty in the
distribution of the observable mass given the true mass causes
ambiguities in the  interpretation of number counts.
 Fig.~\ref{fig:leak} shows the expected mass distribution of clusters above a certain $\Mobs$ given a scatter of $\siglnM=0.25$.  As the observable threshold reaches the exponential tail of
the intrinsic distribution, the excess of upscattered versus downscattered 
clusters can become a significant fraction of
the total. Since at high redshift a fixed $\Mobs$ will be further on the exponential tail,  
even a constant but unknown scatter can introduce a trend in redshift that will degrade the dark energy information in the counts.

\begin{figure}[tb]
\centerline{\epsfxsize=3.4in\epsffile{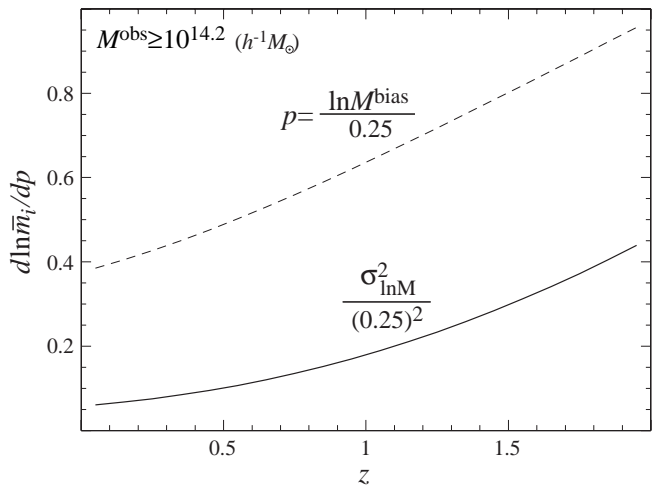}}
\caption{\footnotesize
Fractional sensitivity of the number counts in redshift to the mass variance $\siglnM^{2}$ and
bias $\ln \Mbias$ scaled to $(0.25)^{2}$ and $0.25$ respectively.  The relative importance of the
variance increases with $z$.}
\label{fig:sigsens}
\end{figure}

The relative importance of scatter can be understood by examining the
sensitivity of the counts to the scatter around $\siglnM^2=0$
\begin{align}
\lim_{\siglnM^{2}\rightarrow 0} {\partial \ln \bar m_{i} \over \partial \siglnM^{2}} = - {1\over 2 \bar
n_i}
 {d^{2}\bar n \over d\ln^{2} M} \Big|_{x_{i+1}=0}^{x_{i}=0} \,.
\label{eqn:sig2sens}
 \end{align}
Thus the steepness
 of the mass function around the thresholds in the observable mass determines
the excess due to upscatters.  Note that it is the variance $\siglnM^2$ rather than the 
rms scatter $\siglnM$ that controls the upscattering effect.   For example, since
\begin{align}
\lim_{\siglnM^{2}\rightarrow 0} \left[ 
 {\partial \ln \bar m_{i} \over \partial \siglnM}  =  2\siglnM {\partial \ln \bar m_{i} \over \partial \siglnM^{2}} \right]
 =0 \,,
 \end{align}
the sensitivity to the rms scatter depends on the true value of the scatter and vanishes at
$\siglnM=0$.  
 Conversely from Eqn.~(\ref{eqn:sig2sens}), an observable with say half the scatter would have a quarter of the fractional
 effect on number counts in this limit.
 
On the other hand the sensitivity to the bias is given by
\begin{align}
\lim_{\siglnM^{2}\rightarrow 0} {\partial \ln \bar m_{i} \over \partial \ln \Mbias} = 
{1\over \bar n_i}{d \bar n \over d\ln M} \Big|_{x_{i+1}=0}^{x_{i}=0} \,.
 \end{align}
Thus the relative importance of scatter compared with bias can be estimated through
 the local power law slope of the mass function $d \bar n/d\ln M \propto M^{\alpha}$
 (see Fig.~\ref{fig:slope})
 \begin{align}
 - {1\over 2} 
 {  {d^{2}\bar n / d\ln^{2} M}  \over {d \bar n / d\ln M}  } =-{1\over 2}\alpha(M) \,.
 \label{eqn:alpha}
 \end{align}
Uncertainties in scatter can dominate those of bias for the steep mass function at
high mass or redshift.

 These expectations are borne out  at finite scatter 
  by a direct computation of the number count sensitivity.  Fig.~\ref{fig:sigsens},
 shows the sensitivity of number counts above $\Mobs = 10^{14.2} h^{-1} M_\odot$ 
 in redshift bins of $\Delta z=0.1$ evaluated around $\ln \Mbias=0$ and a finite scatter
 $\siglnM^2=(0.25)^2$.   Both in terms of absolute sensitivity and relative
 sensitivity compared to the bias, the importance of scatter increases with redshift.  
 Uncertainties in the mass variance of $\Delta \siglnM^2 = (0.25)^2$ would
 produce a $\sim 50\%$ uncertainty in the number counts at $z=2$.  For the high $z$ counts
 to provide cosmological information, the scatter must be known to significantly better than
 this level.    
 
 Given that the relative effect of scatter depends on the local slope of the mass function, 
  measuring the
 counts as a function of $\Mobs$ monitors the scatter in the
 mass-observable relation.  Combined with additional information in the sample variance
 of the number counts, an unknown evolution in
  $\Mbias$ and $\siglnM^2$ may be internally calibrated.

\begin{figure}[tb]
\centerline{\epsfxsize=3.4in\epsffile{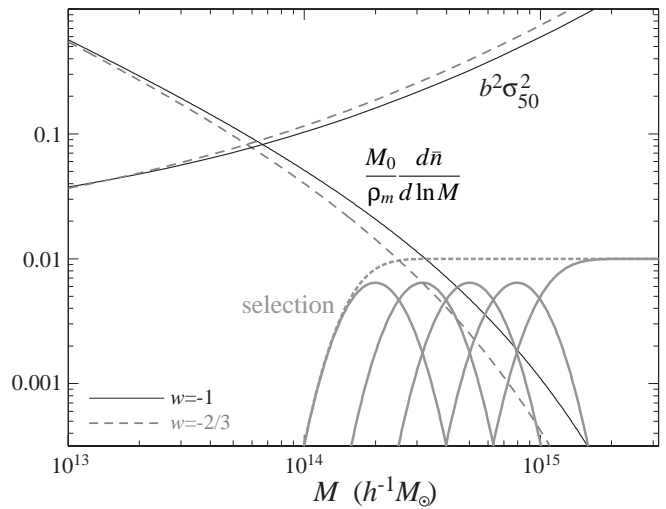}}
\caption{\footnotesize 
Simulations predict the
sample variance of counts, shown for a typical volume ($R=50 h^{-1}$ Mpc, upper curves),
 and the shape of the mass function (lower curves) as a function of mass.
Self-calibration is assisted 
by binning the selection, multiplied here by $0.01$ for clarity and
shown here with $\siglnM=0.25$,
into 5 bins of $\Delta \log_{10} \Mobs =0.2$ (solid lines) as opposed to a single threshold binning of 
 $\Mobs \ge 10^{14.2} h^{-1} M_{\odot}$ (dotted
line).    With only threshold binning, joint changes to the cosmology, mass bias, and
scatter are degenerate with the dark energy equation of state $w$ (long dashed lines).} 
\label{fig:massfn}
\end{figure}

\begin{figure*}[tb]
\centerline{\epsfxsize=7in\epsffile{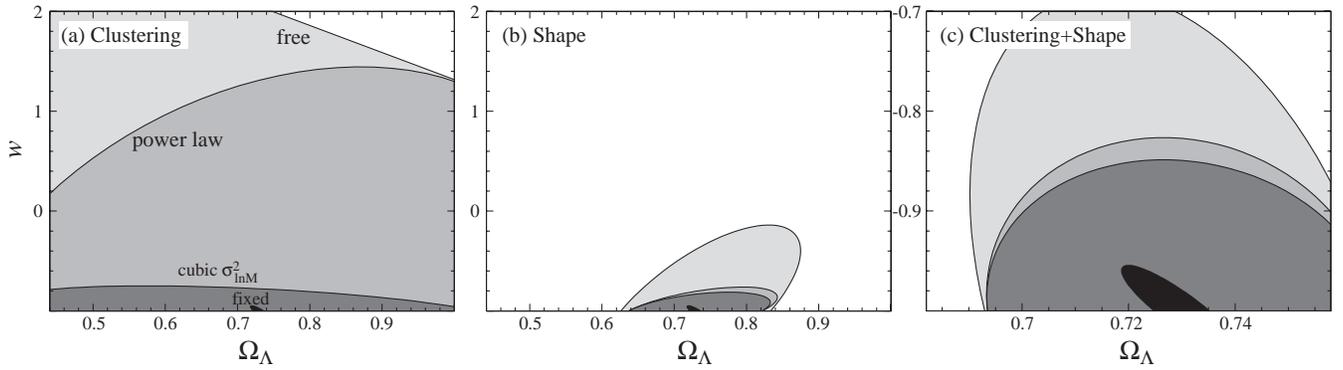}}
\caption{\footnotesize
 Efficacy of self-calibration through
 (a) the clustering
information in the sample
 variance alone, (b) the shape of the counts in 5 mass bins
and (c) both (note the $\times 10$ change in scale).   From outer to inner ellipse, each at the 68\% CL,
the assumptions on the redshift evolution of the
bias and scatter are tightened from a free functional form of 20 parameters each in bins of
$\Delta z=0.1$, a power law evolution in $\Mbias$, an additional cubic form for the
evolution in $\siglnM^{2}$.  Without any form
of self-calibration no relevant dark energy constraints are possible under any of these assumptions.
For reference we show the baseline results of a completely fixed $\Mbias$ and $\siglnM^{2}$ as the
solid innermost ellipse}
\label{fig:ellipse}
\end{figure*}

\section{Self-Calibration}
\label{sec:self}

To assess the impact of uncertainties in the observable-mass distribution, 
we employ the usual Fisher matrix technique.  
For illustrative purposes we take a fiducial cluster survey with specifications
similar to the planned South Pole Telescope (SPT) Survey:
an area of 4000
deg$^{2}$ and a sensitivity corresponding to a constant
$\Mobs_{\rm th} = 10^{14.2} h^{-1} M_{\odot}$.
We  further divide the number counts into bins of redshift $\Delta z=0.1$ from an assumed optical photometric followup out to $z=2$ and 400 angular
cells of 10 deg$^{2}$ for assessment of the sample variance of the counts (see \cite{LimHu04} for
an exploration of these choices).  Finally
to study the efficacy of self-calibration from binning of the observable, we
compare  5 bins of 
$\Delta \log_{10} \Mobs = 0.2$ versus a single bin of
$\Mobs \ge \Mobs_{\rm th}$ (see Fig.~\ref{fig:massfn}).

The Fisher matrix is constructed out of predictions for the number counts and their
covariance.
The mean number counts $m_i$ possess a sample
covariance of \cite{HuKra02}
\begin{eqnarray}
S_{ij} &=&\langle (m_i -\bar m_i)(m_j - \bar m_j)\rangle \nonumber\\
&=&{ b_i \bar m_i b_j \bar m_j  \over V_i V_j} \int{d^3 k \over (2\pi)^3} W_i^*(\bk)W_j(\bk) P(k)\,,
\label{eqn:covariance}
\end{eqnarray}
given a linear power spectrum $P(k)$ and the Fourier transform of the selection window
$W_i(\bx)$.  The pixel index $i$ here runs over unique cells in redshift, angle, and 
observable mass.  
Here $b_i$ is the average bias of the clusters predicted from the distribution 
in Eqn.~(\ref{eqn:binneddensity}) 
\begin{align}
b_i    & =  {1 \over \bar n_i}  \int {d M\over M}  {d \bar n_i \over d\ln M} b(M;z_i)\,,
\end{align}
where we take a fit to simulations of \cite{SheTor99}
\begin{equation}
b(M;z) = 1 + {a_c \delta_c^2/\sigma^2 -1 \over \delta_c} 
         + { 2 p_c \over \delta_c [ 1 + (a \delta_c^2/\sigma^2)^p_c]}
\label{eqn:bias}
\end{equation}
with $a_c=0.75$,  $p_c= 0.3$,  and $\delta_c=1.69$.  That the sample variance, or the clustering
of clusters, is a known
function of mass provides a second means of self-calibration \cite{MajMoh03}.
Note that for a given volume defined by the redshift and solid angle, the mean counts
for different ranges in the observable mass $\Mobs$ are fully correlated.  
Finally the total covariance
matrix is the sample covariance plus shot variance
\begin{eqnarray}
C_{ij} = S_{ij} + \bar m_i \delta_{ij} \,.
\end{eqnarray}

The Fisher matrix quantifies the information in the counts on a set of parameters $p_\alpha$ as
\cite{HolHaiMoh01,LimHu04}
\begin{eqnarray}
F_{\alpha\beta}&=&  \bar{\bf m}^t_{,\alpha} {\bf C}^{-1}
 \bar{\bf m}_{,\beta} 
+ {1\over 2} {\rm Tr} [{\bf C}^{-1} {\bf S}_{,\alpha}
 {\bf C}^{-1} {\bf S}_{,\beta} ]\,,
 \label{eqn:fisher}
\end{eqnarray}
where the first piece represents the information from the mean counts and the second piece the
information from the sample covariance of the counts.   We have here arranged the counts per pixel $i$
into a vector ${\bf m} \equiv (m_1,\ldots,m_{N_{\rm pix}})$ and correspondingly their covariance into
a matrix.
 The Fisher matrix approximates
the covariance matrix of the parameters $C_{\alpha\beta} \approx [{\bf F}^{-1}]_{\alpha\beta}$
such that the marginalized error on a single parameter is $\sigma(p_\alpha) = [{\bf F}^{-1}]_{\alpha\alpha}^{1/2}$.  When considering prior information on parameters of a given $\sigma(p_\alpha)$ we add to the
Fisher matrix a contribution of $\sigma^{-2}(p_\alpha) \delta_{\alpha\beta}$ before inversion.

For the parameters of the Fisher matrix we begin with six cosmological parameters:
the normalization of the initial curvature spectrum 
$\delta_\zeta (=5.07\times 10^{-5})$ at $k=0.05$ Mpc$^{-1}$
(see \cite{HuJai03} for its relationship to the more
traditional $\sigma_8$ normalization), its tilt
$n (=1)$, the baryon density
$\Omega_bh^2 (=0.024)$, the dark matter density
$\Omega_m h^2 (=0.14)$,
and the two dark energy
parameters of interest: its density
$\Omega_{\rm DE} (=0.73)$ relative to critical
and equation of state $w(=-1)$ which we  assume
 to be constant.   
Values in the fiducial cosmology are given in parentheses.  The first 4 parameters
have already been determined at the few to $10\%$ level 
through the CMB \cite{Speetal03} and
we will extrapolate these constraints into the future with priors of
$\sigma(\ln \delta_\zeta)=\sigma(n)=\sigma(\ln\Omega_b h^2)=\sigma(\ln\Omega_m h^2)
=0.01$.  

For the observable-mass distribution we choose a fiducial model of $\ln \Mbias(z_i)=0$ and
$\siglnM^2(z_i) = (0.25)^2$.   The results that follow do not depend on the specific
choice as we have explicitly shown by testing a much smaller fiducial scatter of 
$\siglnM^2(z_i) = (0.05)^2$.   

Given an observable-mass distribution fixed at the fiducial model and the priors
on the other cosmological parameters, the baseline errors on the dark energy 
parameters are $\sigma(\Omega_\Lambda,w)=(0.008,0.03)$.  
The mere presence of scatter in an observable does not necessarily degrade the
cosmological information; in fact for reasonable scatter it actually enhances the information
by effectively lowering the mass threshold at high redshift.

However, cosmological parameter errors are degraded once observable-mass parameters are added 
in a joint fit.    
As the most general case, we take independent $\ln \Mbias(z_{i})$ and $\siglnM^2(z_{i})$ parameters for 
each 
redshift bin for a total of 40 parameters.   As discussed in \S \ref{sec:distribution}, for the
Fisher results to be valid around a fiducial $\siglnM \ll 1$, the mass variance and not its scatter
must be chosen as the parameters.
Because
the evolution in the cluster parameters is expected to be smooth in redshift we
alternatively take a more restrictive power law evolution in the bias $\Mbias$
\begin{equation}
\ln \Mbias(z_i) = A_b + n_b \ln (1+z_i) 
\label{eqn:biaspowerlaw}
\end{equation}
and/or a Taylor expansion of $\siglnM^{2}$ around $z=0$
\begin{equation}
\siglnM^{2}(z_i) = \siglnM^{2} \Big|_{\rm fid} + \sum_{a=0}^{N_\sigma-1} B_a z^a_i
\label{eqn:variancetaylor}
\end{equation}

\begin{figure}[tb]
\centerline{\epsfxsize=3.4in\epsffile{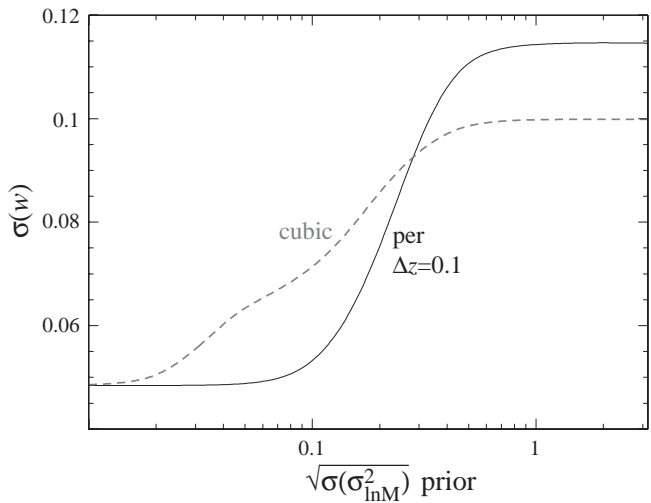}}
\caption{\footnotesize
Prior knowledge of the scatter in the observable-mass relation at the level of
$\sigma( \siglnM^{2})< (0.1)^{2}$ for {\it each} of the 20 redshift bins
can improve dark energy
constraints beyond self-calibration by a factor of 2 (solid line). 
This cumulative effect of independent priors
is compared with the joint effect of
priors on the 4 parameters of a $\sigma(B_{a}) = 2 \sigma(\siglnM^{2})$ of a
cubic $\siglnM^2(z)$ which captures most of the residual uncertainty in Fig.~\ref{fig:ellipse}.  The
latter reflects uncertainties in $\siglnM^{2}$ near $z=1$.}
\label{fig:prior}
\end{figure}

With no self-calibration, i.e. no clustering information from the sample variance
and no binning in $\Mobs$, interesting constraints on the dark energy are
not possible even for the restricted evolutionary forms of 
Eqns.~(\ref{eqn:biaspowerlaw})-(\ref{eqn:variancetaylor}) if $\siglnM^2(z)$ is allowed
to evolve ($N_{\sigma}\ge 2$). 
Even restricting the parameters to a single constant scatter ($N_\sigma=1$) causes
a degradation to $\sigma(\Omega_\Lambda,w)=(0.37,0.24)$.

 As shown in Fig.~\ref{fig:ellipse} adding in the sample
(co)variance information in the Fisher matrix of Eqn.~(\ref{eqn:fisher}) for a single bin
in $\Mobs$ helps but still does not
allow for full self-calibration of an arbitrary evolution in  $\siglnM^2(z)$ even when
$\Mbias$ is restricted  
to power law evolution.  Further restricting the evolution in the mass variance to a cubic
form $N_\sigma=4$ yields $\sigma(\Omega_\Lambda,w)=(0.22,0.17)$
and to a constant form $N_\sigma=1$ yields $\sigma(\Omega_\Lambda,w)=(0.15,0.07)$.

Employing the information contained in the shape of the counts through $\Mobs$
binning allows for a more robust  self-calibration.  In the case of arbitrary evolution for
{\it both} the bias and the scatter $\sigma(\Omega_\Lambda,w)=(0.03,0.21)$.  With
a power law form for the bias  $\sigma(\Omega_\Lambda,w)=(0.02,0.11)$; with
an additional cubic form for the mass variance
  $\sigma(\Omega_\Lambda,w)=(0.02,0.10)$; with
a constant form for the scatter $\sigma(\Omega_\Lambda,w)=(0.02,0.06)$. 

External priors on the observable-mass distribution from simulations and cross-calibration
of observables can further improve on self-calibration.  
Cross-calibration of cluster observables may involve a subsample of clusters which
have detailed mass modeling from lensing or X-ray temperature and surface brightness
profiles assuming hydrostatic equilibrium  \cite{MajMoh03}.
 In Fig.~\ref{fig:prior} we explore
the effect of independent priors on the 20 $\siglnM^2(z_i)$ parameters
in the power law $\Mbias$ context.  Priors of the level of $\sigma(  \siglnM^2 )=(0.1)^2$ would suffice
to improve $\sigma(w)$ by a factor of 2.  Note that the potential
further improvement in $w$ errors
comes from the ability to change the scatter smoothly from $z=0$ to  $z\sim 1$.  
Since we take the priors to be independent, their cumulative effect
implicitly poses a much more stringent constraint on the possible
smooth evolution of $\siglnM^2$  than any one individual prior.

To better quantify the implications of the joint prior, note that  the
self-calibration errors on the cubic form in Fig.~\ref{fig:ellipse}c
 nearly coincide with the fully arbitrary form.
Taking independent 
priors on the 4 $B_a$ parameters of $\sigma(B_a)= 2 \sigma(\siglnM^2)$ to reflect the assumed
uncertainty around $z=1$ yields the dashed curve in Fig.~\ref{fig:prior}. The full improvement requires $\siglnM^2$ priors at the $(0.02)^2$ level and
a minimum of $(0.1)^2$ for substantial improvements. If these priors are to come from
mass modeling of observables then a {\it fair} sample of more than $\sim 100$ clusters
at $z\sim 1$ with accurate masses will be required.  Accurate masses will be
difficult to obtain at the low threshold of $10^{14.2} h^{{-1}} M_{\odot}$ employed here.

Priors on $\Mbias$ can also improve constraints.   For a completely fixed $\Mbias$,
errors for an arbitrary evolution in $\siglnM^2$ are $\sigma(\Omega_\Lambda,w)=(0.01,0.06)$.
Conversely for a completely fixed $\siglnM^2$, errors for an arbitrary evolution of 
$\Mbias$ are $\sigma(\Omega_\Lambda,w)=(0.02,0.13)$.    

Finally to assess the possible impact of unknown trends in the mass bias and variance
with mass at fixed redshift, we 
limit the $\Mobs$ bins to 2 separated by $\Delta \log_{10}\Mobs=0.2$ from the threshold.
In this case errors for arbitrary evolution degrade slightly to 
$\sigma(\Omega_\Lambda,w)=(0.03,0.26)$ and with power law mass bias and cubic
variance to $\sigma(\Omega_\Lambda,w)=(0.03,0.11)$.  Thus most of the information from
self-calibration comes from the small range in masses around the threshold reflecting
the steepness of the mass function.  
The mass bias and mass variance need only be constant or slowly varying in a known way
in mass across a range in masses comparable to the expected scatter for self-calibration
to be effective.   In any case, bins at higher masses also monitor the validity of this assumption
in practice.

\smallskip

\section{Discussion}
\label{sec:discussion}

The exponential sensitivity of number counts to the cluster mass requires a calibration of
the whole observable-mass distribution before cosmological information on 
the dark energy can be 
extracted.
We have shown that even in the case of an unknown arbitrary evolution in the mass bias and scatter of
a Gaussian distribution there is enough information in the ratio
 of the counts in bins of the 
observable mass and their sample 
variance to calibrate the distribution and provide interesting constraints
on the dark energy.  

For the more realistic case of an unknown power law evolution in the
mass bias $\Mbias$, the forecasted
errors for the fiducial SPT-like survey are $\sigma(\Omega_\Lambda,w) = (0.02,0.11)$.
To further improve on these constraints, external constraints on the mass variance would need
to achieve an accuracy of $\sigma( \siglnM^2 ) < (0.1)^2 =0.01$ on a possible evolution of
the mass variance from $0\le z \le 1$.   Note that this result is robust
to the assumed true value of the scatter when quoted as a constraint on the mass variance
and not the rms scatter.

However, 
self-calibration is not a replacement for simulated catalogues,
cross-calibration techniques from so-called direct mass measurements 
\cite{MajMoh03}, and monitoring scatter in observable-observable scaling relations.
It is instead an internal consistency check on their assumptions and the simplifying
assumptions in this study.
We have assumed that the observable-mass distribution
is a Gaussian in $\ln M$ and that its parameters depend in a known way on mass at a given redshift for at least a range in masses that is greater than the expected scatter.
Furthermore, for low mass
clusters detected optically (e.g. \cite{GlaYee05}) or through lensing (e.g. \cite{Witetal03}), 
the assumption of a one-to-one mapping of
objects identified by mass to those identified by the observable breaks down 
since confusion and projection will cause
many small mass objects in a given redshift range
to be associated with a single object in the observable (see e.g. \cite{Kimetal02,HenSpe04}).

Without such simplifying assumptions, true
self-calibration is impossible (see e.g. \cite{Hu03a}).
 Still, the ideas underlying self-calibration will be useful in 
revealing violations
of the assumed form  of the
distribution of cluster observables given the cluster 
masses and prevent misinterpretation of the data.   
\smallskip

{
\noindent{\it Acknowledgments:}  We thank D. Hogg,
G. Holder,  A. Kravtsov, J. Mohr, A. Schulz, T. McKay,  A. Vikhlinin, J. Weller, M. White
for useful discussions.  We also thank the organizer, A. Evrard, and the participants of
the Kona Cluster meeting.  
This work was supported by the DOE, the Packard Foundation and CNPq;
it was carried out  at  the KICP under NSF PHY-0114422.  }
\smallskip

%\bibliography{LimHu05}

\begin{thebibliography}{20}
\expandafter\ifx\csname natexlab\endcsname\relax\def\natexlab#1{#1}\fi
\expandafter\ifx\csname bibnamefont\endcsname\relax
  \def\bibnamefont#1{#1}\fi
\expandafter\ifx\csname bibfnamefont\endcsname\relax
  \def\bibfnamefont#1{#1}\fi
\expandafter\ifx\csname citenamefont\endcsname\relax
  \def\citenamefont#1{#1}\fi
\expandafter\ifx\csname url\endcsname\relax
  \def\url#1{\texttt{#1}}\fi
\expandafter\ifx\csname urlprefix\endcsname\relax\def\urlprefix{URL }\fi
\providecommand{\bibinfo}[2]{#2}
\providecommand{\eprint}[2][]{\url{#2}}

\bibitem[{\citenamefont{{Ikebe} et~al.}(2002)\citenamefont{{Ikebe}, {Reiprich},
  {Boehringer}, {Tanaka}, and {Kitayama}}}]{Ikeetal01}
\bibinfo{author}{\bibfnamefont{Y.}~\bibnamefont{{Ikebe}}},
  \bibinfo{author}{\bibfnamefont{T.}~\bibnamefont{{Reiprich}}},
  \bibinfo{author}{\bibfnamefont{H.}~\bibnamefont{{Boehringer}}},
  \bibinfo{author}{\bibfnamefont{Y.}~\bibnamefont{{Tanaka}}}, \bibnamefont{and}
  \bibinfo{author}{\bibfnamefont{T.}~\bibnamefont{{Kitayama}}},
  \bibinfo{journal}{Astron. Astrophys.} \textbf{\bibinfo{volume}{383}},
  \bibinfo{pages}{773} (\bibinfo{year}{2002}), \eprint{astro-ph/0112315}.

\bibitem[{\citenamefont{{Haiman} et~al.}(2001)\citenamefont{{Haiman}, {Mohr},
  and {Holder}}}]{HaiMohHol01}
\bibinfo{author}{\bibfnamefont{Z.}~\bibnamefont{{Haiman}}},
  \bibinfo{author}{\bibfnamefont{J.}~\bibnamefont{{Mohr}}}, \bibnamefont{and}
  \bibinfo{author}{\bibfnamefont{G.}~\bibnamefont{{Holder}}},
  \bibinfo{journal}{\apj} \textbf{\bibinfo{volume}{553}}, \bibinfo{pages}{545}
  (\bibinfo{year}{2001}).

\bibitem[{\citenamefont{{Hu} and {Kravtsov}}(2003)}]{HuKra02}
\bibinfo{author}{\bibfnamefont{W.}~\bibnamefont{{Hu}}} \bibnamefont{and}
  \bibinfo{author}{\bibfnamefont{A.}~\bibnamefont{{Kravtsov}}},
  \bibinfo{journal}{\apj} \textbf{\bibinfo{volume}{584}}, \bibinfo{pages}{702}
  (\bibinfo{year}{2003}), \eprint{astro-ph/0203169}.

\bibitem[{\citenamefont{{Wang} et~al.}(2004)\citenamefont{{Wang}, {Khoury},
  {Haiman}, and {May}}}]{WanKhoHaiMay04}
\bibinfo{author}{\bibfnamefont{S.}~\bibnamefont{{Wang}}},
  \bibinfo{author}{\bibfnamefont{J.}~\bibnamefont{{Khoury}}},
  \bibinfo{author}{\bibfnamefont{Z.}~\bibnamefont{{Haiman}}}, \bibnamefont{and}
  \bibinfo{author}{\bibfnamefont{M.}~\bibnamefont{{May}}},
  \bibinfo{journal}{\prd} \textbf{\bibinfo{volume}{70}},
  \bibinfo{pages}{123008} (\bibinfo{year}{2004}).

\bibitem[{\citenamefont{{Eddington}}(1913)}]{Edd13}
\bibinfo{author}{\bibfnamefont{A.}~\bibnamefont{{Eddington}}},
  \bibinfo{journal}{\MNRAS} \textbf{\bibinfo{volume}{73}}, \bibinfo{pages}{359}
  (\bibinfo{year}{1913}).

\bibitem[{\citenamefont{{Holder} et~al.}(2000)\citenamefont{{Holder}, {Mohr},
  {Carlstrom}, {Evrard}, and {Leich}}}]{HolMohCarEvrLei00}
\bibinfo{author}{\bibfnamefont{G.}~\bibnamefont{{Holder}}},
  \bibinfo{author}{\bibfnamefont{J.}~\bibnamefont{{Mohr}}},
  \bibinfo{author}{\bibfnamefont{J.}~\bibnamefont{{Carlstrom}}},
  \bibinfo{author}{\bibfnamefont{A.}~\bibnamefont{{Evrard}}}, \bibnamefont{and}
  \bibinfo{author}{\bibfnamefont{E.}~\bibnamefont{{Leich}}},
  \bibinfo{journal}{\apj} \textbf{\bibinfo{volume}{544}}, \bibinfo{pages}{629}
  (\bibinfo{year}{2000}), \eprint{astro-ph/9912364}.

\bibitem[{\citenamefont{{Battye} and {Weller}}(2003)}]{BatWel03}
\bibinfo{author}{\bibfnamefont{R.}~\bibnamefont{{Battye}}} \bibnamefont{and}
  \bibinfo{author}{\bibfnamefont{J.}~\bibnamefont{{Weller}}},
  \bibinfo{journal}{\prd} \textbf{\bibinfo{volume}{68}},
  \bibinfo{pages}{083506} (\bibinfo{year}{2003}).

\bibitem[{\citenamefont{{Levine} et~al.}(2002)\citenamefont{{Levine}, {Schulz},
  and {White}}}]{LevSchWhi02}
\bibinfo{author}{\bibfnamefont{E.}~\bibnamefont{{Levine}}},
  \bibinfo{author}{\bibfnamefont{A.}~\bibnamefont{{Schulz}}}, \bibnamefont{and}
  \bibinfo{author}{\bibfnamefont{M.}~\bibnamefont{{White}}},
  \bibinfo{journal}{\apj} \textbf{\bibinfo{volume}{577}}, \bibinfo{pages}{569}
  (\bibinfo{year}{2002}), \eprint{astro-ph/0204273}.

\bibitem[{\citenamefont{{Hu}}(2003)}]{Hu03a}
\bibinfo{author}{\bibfnamefont{W.}~\bibnamefont{{Hu}}}, \bibinfo{journal}{\prd}
  \textbf{\bibinfo{volume}{67}}, \bibinfo{pages}{081304}
  (\bibinfo{year}{2003}), \eprint{astro-ph/0301416}.

\bibitem[{\citenamefont{{Majumdar} and {Mohr}}(2003)}]{MajMoh03}
\bibinfo{author}{\bibfnamefont{S.}~\bibnamefont{{Majumdar}}} \bibnamefont{and}
  \bibinfo{author}{\bibfnamefont{J.}~\bibnamefont{{Mohr}}},
  \bibinfo{journal}{\apj} \textbf{\bibinfo{volume}{585}}, \bibinfo{pages}{603}
  (\bibinfo{year}{2003}), \eprint{astro-ph/0208002}.

\bibitem[{\citenamefont{{Lima} and {Hu}}(2004)}]{LimHu04}
\bibinfo{author}{\bibfnamefont{M.}~\bibnamefont{{Lima}}} \bibnamefont{and}
  \bibinfo{author}{\bibfnamefont{W.}~\bibnamefont{{Hu}}},
  \bibinfo{journal}{\prd} \textbf{\bibinfo{volume}{70}},
  \bibinfo{pages}{043504} (\bibinfo{year}{2004}), \eprint{astro-ph/0401559}.

\bibitem[{\citenamefont{{Jenkins} et~al.}(2001)}]{Jenetal01}
\bibinfo{author}{\bibfnamefont{A.}~\bibnamefont{{Jenkins}}}
  \bibnamefont{et~al.}, \bibinfo{journal}{\MNRAS}
  \textbf{\bibinfo{volume}{321}}, \bibinfo{pages}{372} (\bibinfo{year}{2001}).

\bibitem[{\citenamefont{{Sheth} and {Tormen}}(1999)}]{SheTor99}
\bibinfo{author}{\bibfnamefont{R.}~\bibnamefont{{Sheth}}} \bibnamefont{and}
  \bibinfo{author}{\bibfnamefont{B.}~\bibnamefont{{Tormen}}},
  \bibinfo{journal}{\MNRAS} \textbf{\bibinfo{volume}{308}},
  \bibinfo{pages}{119} (\bibinfo{year}{1999}).

\bibitem[{\citenamefont{{Holder} et~al.}(2001)\citenamefont{{Holder}, {Haiman},
  and {Mohr}}}]{HolHaiMoh01}
\bibinfo{author}{\bibfnamefont{G.}~\bibnamefont{{Holder}}},
  \bibinfo{author}{\bibfnamefont{Z.}~\bibnamefont{{Haiman}}}, \bibnamefont{and}
  \bibinfo{author}{\bibfnamefont{J.}~\bibnamefont{{Mohr}}},
  \bibinfo{journal}{\apj Lett.} \textbf{\bibinfo{volume}{560}},
  \bibinfo{pages}{111} (\bibinfo{year}{2001}).

\bibitem[{\citenamefont{{Hu} and {Jain}}(2004)}]{HuJai03}
\bibinfo{author}{\bibfnamefont{W.}~\bibnamefont{{Hu}}} \bibnamefont{and}
  \bibinfo{author}{\bibfnamefont{B.}~\bibnamefont{{Jain}}},
  \bibinfo{journal}{\prd} \textbf{\bibinfo{volume}{70}},
  \bibinfo{pages}{043009} (\bibinfo{year}{2004}), \eprint{astro-ph/0312395}.

\bibitem[{\citenamefont{{Spergel} et~al.}(2003)}]{Speetal03}
\bibinfo{author}{\bibfnamefont{D.}~\bibnamefont{{Spergel}}}
  \bibnamefont{et~al.}, \bibinfo{journal}{\apj Sup.}
  \textbf{\bibinfo{volume}{148}}, \bibinfo{pages}{175} (\bibinfo{year}{2003}),
  \eprint{astro-ph/0302209}.

\bibitem[{\citenamefont{{Gladders} and {Yee}}(2005)}]{GlaYee05}
\bibinfo{author}{\bibfnamefont{M.}~\bibnamefont{{Gladders}}} \bibnamefont{and}
  \bibinfo{author}{\bibfnamefont{H.}~\bibnamefont{{Yee}}},
  \bibinfo{journal}{\apj Supp.} \textbf{\bibinfo{volume}{157}},
  \bibinfo{pages}{1} (\bibinfo{year}{2005}).

\bibitem[{\citenamefont{{Wittman} et~al.}(2003)}]{Witetal03}
\bibinfo{author}{\bibfnamefont{D.}~\bibnamefont{{Wittman}}}
  \bibnamefont{et~al.}, \bibinfo{journal}{\apj} \textbf{\bibinfo{volume}{597}},
  \bibinfo{pages}{218} (\bibinfo{year}{2003}).

\bibitem[{\citenamefont{{Kim} et~al.}(2002)}]{Kimetal02}
\bibinfo{author}{\bibfnamefont{R.}~\bibnamefont{{Kim}}} \bibnamefont{et~al.},
  \bibinfo{journal}{Astron. J.} \textbf{\bibinfo{volume}{310}},
  \bibinfo{pages}{31} (\bibinfo{year}{2002}).

\bibitem[{\citenamefont{{Hennawi} and {Spergel}}(2005)}]{HenSpe04}
\bibinfo{author}{\bibfnamefont{J.}~\bibnamefont{{Hennawi}}} \bibnamefont{and}
  \bibinfo{author}{\bibfnamefont{D.}~\bibnamefont{{Spergel}}},
  \bibinfo{journal}{\apj} \textbf{\bibinfo{volume}{\rm in press}},
  \bibinfo{pages}{astro} (\bibinfo{year}{2005}).

\end{thebibliography}

\vfill

\end{document}